\documentclass{article}

\usepackage{arxiv}

\usepackage[utf8]{inputenc} 
\usepackage[T1]{fontenc}    
\usepackage{hyperref}       
\usepackage{url}            
\usepackage{booktabs}       
\usepackage{nicefrac}       
\usepackage{microtype}      
\usepackage{lipsum}		
\usepackage{graphicx}
\usepackage[numbers,sort&compress]{natbib}
\usepackage{doi}
\usepackage{xcolor}
\usepackage{stix}
\usepackage{amsmath}
\usepackage{bm}

\definecolor{forestgreen}{rgb}{0.1328125, 0.54296875, 0.1328125}
\definecolor{navy}{rgb}{0.0, 0.0, 0.5}
\definecolor{red}{rgb}{1.0, 0.0, 0.0}
\definecolor{darkslategray}{rgb}{0.1845703125, 0.3095703125, 0.3095703125}
\definecolor{dodgerblue}{rgb}{0.1171875, 0.5625, 1.0}
\definecolor{orange}{rgb}{1.0, 0.647, 0.0}

\title{Shape Matters: Detecting Vertebral Fractures Using Differentiable Point-Based Shape Decoding}

\author{ \href{https://orcid.org/0009-0002-3669-6751}{\includegraphics[scale=0.06]{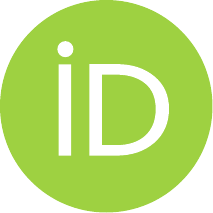}\hspace{1mm}Hellena Hempe} \\
	Institute of Medical Informatics\\
	  University of Luebeck\\
	Luebeck, Germany \\
	\texttt{hellena.hempe@uni-luebeck.de} \\
	\And
	\href{https://orcid.org/0000-0001-7824-5735}{\includegraphics[scale=0.06]{orcid.pdf}\hspace{1mm}Alexander Bigalke\thanks{Alexander Bigalke is also affiliated with Dräger, Drägerwerk AG \& Co. KGaA}} \\
	Institute of Medical Informatics\\
	  University of Luebeck\\
	Luebeck, Germany \\
	\texttt{alexander.bigalke@uni-luebeck.de} \\
    \And
	\href{https://orcid.org/0000-0002-7489-1972}{\includegraphics[scale=0.06]{orcid.pdf}\hspace{1mm}Mattias P.~Heinrich} \\
	Institute of Medical Informatics\\
	  University of Luebeck\\
	Luebeck, Germany \\
	\texttt{mattias.heinrich@uni-luebeck.de} \\
}

\renewcommand{\shorttitle}{\textit{Shape Matters: Detecting Vertebral Fractures}}

\hypersetup{
pdftitle={shape_matters_arxiv},
pdfsubject={q-bio.NC, q-bio.QM},
pdfauthor={Hellena Hempe, Alexander Bigalke, Mattias P. Heinrich},
pdfkeywords={First keyword, Second keyword, More},
}

\begin{document}
\maketitle

\begin{abstract}
	Degenerative spinal pathologies are highly prevalent among the elderly population. Timely diagnosis of osteoporotic fractures and other degenerative deformities facilitates proactive measures to mitigate the risk of severe back pain and disability. 
\textbf{Objective: }In this study, we specifically explore the use of shape auto-encoders for vertebrae, taking advantage of advancements in automated multi-label segmentation and the availability of large datasets for unsupervised learning. 
\textbf{Methods: }Our shape auto-encoders are trained on a large set of vertebrae surface patches, leveraging the vast amount of available data for vertebra segmentation. This addresses the label scarcity problem faced when learning shape information of vertebrae from image intensities. Based on the learned shape features we train an MLP to detect vertebral body fractures.
\textbf{Results: }Using segmentation masks that were automatically generated using the TotalSegmentator, our proposed method achieves an AUC of 0.901 on the VerSe19 testset. This outperforms image-based and surface-based end-to-end trained models. Additionally, our results demonstrate that pre-training the models in an unsupervised manner enhances geometric methods like PointNet and DGCNN.
\textbf{Conclusion: }Our findings emphasise the advantages of explicitly learning shape features for diagnosing osteoporotic vertebrae fractures. This approach improves the reliability of classification results and reduces the need for annotated labels.
\textbf{Significance: }This study provides novel insights into the effectiveness of various encoder-decoder models for shape analysis of vertebrae and proposes a new decoder architecture: the point-based shape decoder.
The source code will be made available on GitHub: \href{https://github.com/multimodallearning/shape_matters}{https://github.com/multimodallearning/shape\_matters}
\end{abstract}


\section{Introduction}

\label{sec:introduction}

Degenerative pathology of the spine such as osteoporotic fractures of vertebral bodies, spinal canal stenosis, spondylolisthesis and other degenerative deformations of vertebrae, are a common healthcare risk in the elderly population \citep{ballane_worldwide_2017}. Early detection of degenerative diseases may enable preventative measures to reduce the risk of chronic severe back pain and disability \citep{papaioannou_diagnosis_2002}. 
In recent years, deep learning has emerged as a powerful tool for many medical applications. Advances in deep learning in computer-aided diagnosis have paved the way for more timely interventions and improved patient outcomes.
In this study, we specifically focus on the intersection of deep learning and the diagnosis of vertebral body fractures.

So far, most deep learning-based approaches for classification of vertebral body fractures are trained on intensity images in an end-to-end manner.
Even after the VerSe dataset/benchmark for segmentation and localisation of vertebrae was introduced in '19 and '20 \citep{liebl_computed_2021,loffler_vertebral_2020, sekuboyina_verse_2021}, most introduced methods rely on the localisation of vertebrae instead of leveraging the available segmentation masks \citep{nicolaes_detection_2020, yilmaz_automated_2021, zakharov_interpretable_2022}. The amount of available ground truth segmentation masks of vertebrae was further increased by the TotalSegmentator dataset \citep{ts_rsna,isensee_nnu-net_2021}. To date, only few recently proposed methods \citep{husseini_conditioned_2020, sekuboyina_probabilistic_2019} employ the information provided by segmentation masks in diagnostic downstream tasks.

We identified three relevant challenges that are not sufficiently addressed by existing methods and have so far prevented a wider clinical adoption of vertebral CAD.
a) Image-based classifiers are prone to deterioration under domain shifts, i.e. they are limited in their adaptability for variations of the image intensity, scanner acquisition settings and other shifts in the data distribution. Furthermore, 3D CNNs trained in a fully-supervised setting tend to overfit and require a larger amount of labelled data of sufficient quality. 
b) Surface-based geometric learning models \citep{qi2017pointnet,wang2019dynamic} have so far been less expressive than 3D CNNs and may not achieve the required accuracy on limited labelled data. 
c) Shape encoder-decoder architectures \citep{husseini_conditioned_2020, sekuboyina_probabilistic_2019} may help to overcome the label scarcity by shifting a larger proportion of the training to an unsupervised setting for which even healthy spines can be employed, but they may still fail to learn representative latent space representations, due to an over-pronounced weight of the decoder (that is later discarded for the classification task).   

As a remedy, we make the following suggestions:
a) Instead of training image-based classifiers we propose to leverage information from a preceding segmentation model and directly operate on shape information of vertebrae (surface contour) for the classification task. This allows trained deep learning models to be independent of shifts in image intensities and moves the demand for labelled data away from the classification task.
b) Leveraging the vast amount of available segmentation masks, unsupervised learning can help geometric learning models overcome problems related to limited labelled data.
c) To ensure a more representative latent space representation, we introduce a novel decoder architecture that ensures that most learned parameters are located in the encoder model and thus relevant for the classification task.

\textbf{Contributions: } 1) We believe that the effectiveness of recently proposed AE models is limited due to the sub-optimal design of encoder-decoder components. Therefore, we perform an in-depth analysis of the effectiveness of various AE architectures for the diagnosis of vertebral body fractures.
2) For this purpose, we develop a modular encoder-decoder framework which allows arbitrary combinations of point- and image-based encoder and decoder architectures.
3) To address the problem of over-pronounced weights in the decoder, we designed a novel point-based shape decoder model that employs a differentiable sampling step to bridge the gap between point- and voxel-based solutions.
4) By performing extensive experiments to analyse various combinations of encoder and decoder architectures we verified that the detection of vertebral fractures, which by definition is a geometric problem, is better solved using shape-based methods compared to image-intensity-based.
5) The results of our experiments furthermore demonstrate the particular advantages of employing our novel point-based shape decoder compared to other models.

\textbf{Outline:} In the subsequent section \ref{sec:relatedworks} we provide a comprehensive review of related literature, emphasising the existing disparity between image- and point-based analyses. Following this, our comprehensive shape-based classification framework \ref{subsec:buildingblocks}, incorporating the proposed point-based shape decoder \ref{subsec:pbshapedecoder}, is meticulously detailed in the methods section \ref{sec:methods}. The experiments and results section \ref{sec:experimentsresults} elaborates on our setup, encompassing data particulars and implementation details \ref{subsec:datasetdetails}. Afterwards, the results of our experiments are presented in \ref{subsec:fxpipeline} and \ref{subsec:dataholdexperiment}, followed by a thorough discussion of the outcomes \ref{sec:discussion} and a concise summary of our findings \ref{sec:conclusion}.

\section{Related Works}
\label{sec:relatedworks}
Radiologists commonly assess osteoporotic fractures using the Genant score \citep{genant_vertebral_1993}. The Genant Score is a semiquantitative measurement by which a fracture is defined by the height reduction of the vertebral body (in lateral X-Ray) and categorised into 0-healthy, 1-mildly fractured, 2-moderately fractured and 3-severely fractured. In 3D-CT scans this is a bit more complicated as an optimal slice for measuring the height difference needs to be determined. However, automated detection of vertebral fractures is often performed on CT scans especially on quantitative CT scans for measurements of bone mineral density which is directly linked to osteoporosis \citep{loffler_automatic_2021}.

When categorising recently proposed methods for computer-aided detection of vertebral compression fractures, we can observe that most recently proposed methods operate directly on the CT intensity data e.g. \citep{nicolaes_detection_2020, yilmaz_automated_2021, husseini_grading_2020}. The method proposed by Nicolaes et al. \citep{nicolaes_detection_2020} suggests a two-step approach which detects vertebral fractures in CT scans using a voxel-based CNN to perform voxel-level classification on patches containing vertebrae.
Yilmaz et al. \citep{yilmaz_automated_2021} present a pipeline in which the vertebrae are firstly detected using a hierarchical neural network and secondly apply a CNN on the identified patches that are extracted from the original CT scan. 
The Grading Loss as proposed by Husseini et al. \citep{husseini_grading_2020}, addresses the problem of severe data imbalance and minor differences in appearance between healthy and fractured vertebrae in CT scans by exploring a representation learning-based approach that was inspired by Genant grades.
These methods do not exploit the shape information provided by the segmentation mask. Consequently, the shape of the vertebrae and the associated fracture status need to be learned implicitly, which leads to a complication of the learning process.

As a remedy, two related previous works aim to explicitly use the shape of vertebrae. Firstly, the Spine-VAE \citep{husseini_conditioned_2020} employs the masked image data of vertebrae as input to a conditioned VAE to capture the large variations of healthy and fractured vertebra shapes. The conditioning is performed by concatenating information about the label of the corresponding to the vertebra (T1-L5 split into 5 groups) as one-hot-encoded vector. After training the VAE until the reconstruction loss converges, the encoder model is further in conjunction with a Multi-Layer Perceptron to classify vertebral body fractures.
Secondly, the purely geometric approach proposed by Sekuboyina et al. \citep{sekuboyina_probabilistic_2019} employs a surface point-based probabilistic AE to learn shape representations of vertebrae. The task of detecting vertebral fractures is then treated as out-of-distribution detection task by computing the reconstruction error based on a model that was only trained on healthy vertebrae subjects.
Both methods rely on accurate segmentation masks of vertebrae during test-time and do not involve unsupervised pre-training on a larger separate dataset to learn shape features.

In many computer vision applications, geometric deep learning methods play a crucial role in extracting meaningful features from 3D data by capturing spatial relationships for shape analysis. Notably, two state-of-the-art techniques in this domain are PointNet \citep{qi2017pointnet} and DGCNN (Dynamic Graph Convolutional Neural Network)\citep{zhang2018end}. PointNet stands out for its ability to process unstructured point clouds directly, showcasing robust performance in tasks like shape recognition and segmentation. DGCNN, on the other hand, utilises dynamic graph structures to capture both local and global geometric relationships within point cloud data, enhancing the model's capacity to discern intricate patterns. These geometric deep learning approaches have significantly advanced the accuracy and efficiency of medical image analysis, particularly in scenarios where a nuanced understanding of spatial relationships and complex structures within volumetric data is essential.

\begin{figure*}[!t]
\centering
\includegraphics[width=6.6in]{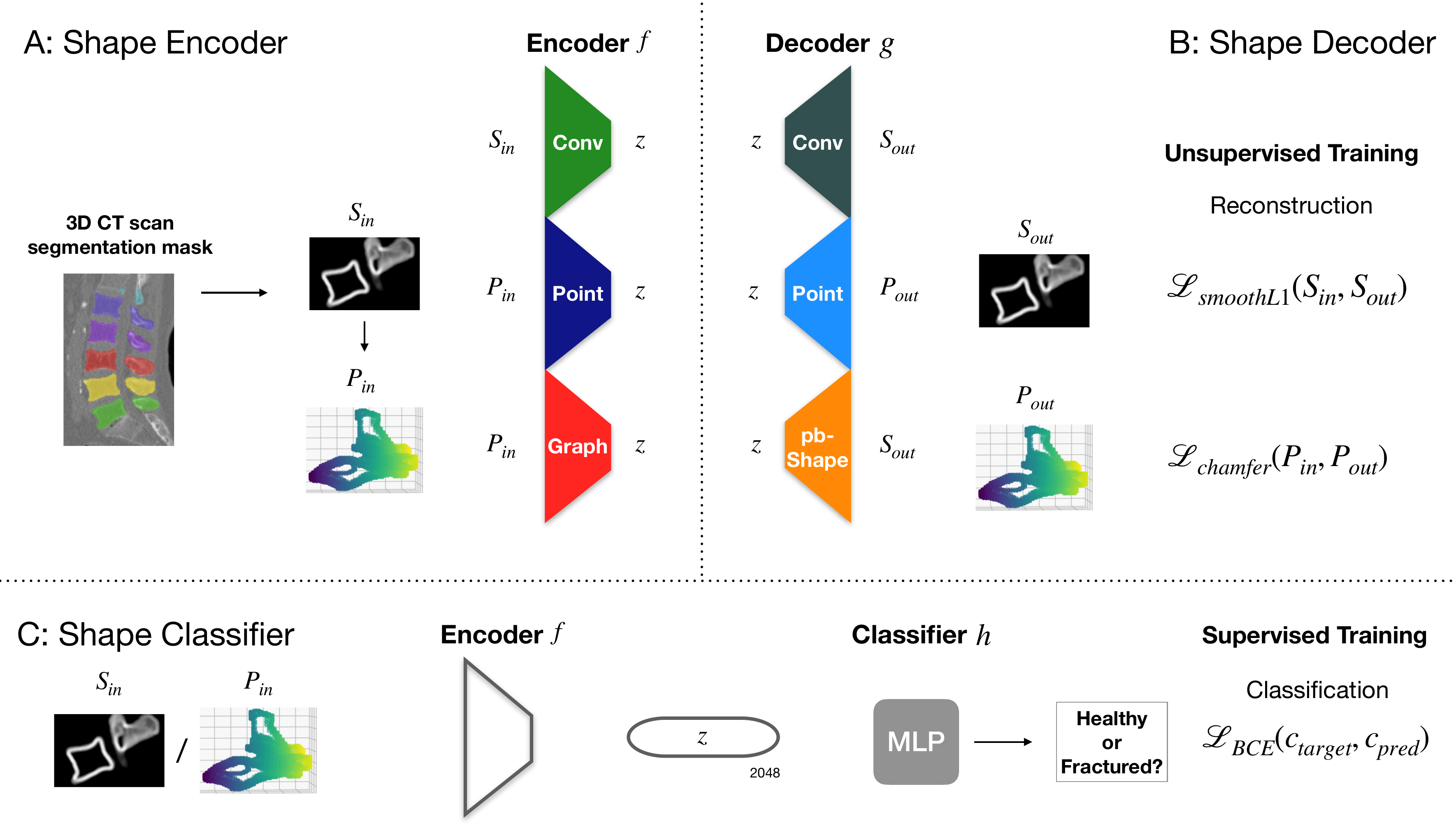}
\caption{To determine the most suitable architecture for our task we employ combinations of several encoder-decoder architectures including traditional convolutional methods and geometric methods. The AEs are trained to reconstruct either a point-cloud representation or a volumetric surface representation of vertebrae, which are derived from the previously computed segmentation mask. As \textbf{Shape encoder (A)} we employ a convolutional method, as well as a point-based and a graph-based method to predict the embedding $\bm{z}$. As \textbf{Shape decoder (B)} we employ a convolutional method as well as a point-based method and propose a novel point-based shape decoder. The \textbf{Shape classifier (C)} is then trained separately on the embedding $\bm{z}$ for each encoder-decoder combination using the same MLP-model. Note that only the weights of the MLP are trained in a supervised manner, whereas the weights of the encoder are fixed.}
\label{fig_method_a}
\end{figure*}

Geometric learning approaches are already being applied in medical imaging tasks, e.g. assessment for spinal curvature in X-Ray \citep{huo_joint_2021} and detection of pedicles for spinal surgery in CT scans \citep{burgin_robust_2023}. 
However, detecting vertebral fractures in 3D CT scans allows for automated opportunistic screening in CT scans that were taken for other purposes \citep{engelke_opportunistic_2023}. Furthermore, the required landmarks for the measurement of vertebral height in 2D lie on the surface of the vertebral body in 3D, which is why we believe that geometric learning approaches are beneficial for learning the important shape features from the contour of vertebrae.

\section{Methods}
\label{sec:methods}

To investigate the potential of shape features for detecting vertebral fractures, we experiment with several encoder-decoder architectures and deploy their latent space vector for classification. In particular, we compare combinations of encoder and decoder trained to reconstruct shape features of vertebrae based on the surface of their respective segmentation mask. The training of our AE architecture comparison pipeline is split into two stages. In the first stage, we train our AE models in an unsupervised manner on a large-scale dataset without classification labels or a specific high occurrence of spine diseases to generate meaningful shape features in the latent space. During the second stage, we employ the generated shape features of the encoder (freezing the encoder layers) and train an MLP for detection of fractures based on these shape features on a smaller labelled dataset.

The problem setup can be defined as follows:
Based on multi-label segmentation masks of vertebrae performed on a CT Scan, extract patches  $\mathcal{S}_{in}\in\mathbb{R}^{D\times H\times W}$ of the segmentation mask contour that is computed using an edge-filter. For geometric learning methods, we extract a point cloud representation $\mathcal{P}_{in}\in\mathbb{R}^{3\times N}$ of $N$ 3D points by employing a threshold on the grid. Firstly, train an AE-model comprised of a shape encoder $f$ and a shape decoder $g$ on an unsupervised reconstruction task by minimising the smoothL1-Loss of the Chamfer-Loss between the input ${S}_{in}$ or ${P}_{in}$ and the prediction ${S}_{out}$ or ${P}_{out}$. Secondly, using the latent space vector $\bm{z}\in\mathbb{R}^M$ computed by the shape encoder $f$ as input, train an MLP as classifier $h$ to predict the fracture status (healthy or fractured) by minimising the Cross-Entropy loss function between the target class ${c}_{target}$ and the predicted class ${c}_{pred}$.
As encoders we employ 3 approaches, a simple convolutional-encoder ${f}_{conv}$, a point-encoder ${f}_{point}$ based on the PointNet architecture \citep{qi2017pointnet} and a Graph encoder ${f}_{graph}$ based on the DGCNN \citep{zhang2018end}.
As decoders, we employ 3 approaches, a convolutional decoder ${g}_{conv}$, our a point-decoder ${g}_{point}$ and a novel point-based shape decoder ${g}_{pbShape}$
The main idea of the general framework for the employed image- or point-based vertebrae auto-encoder (AE) framework is depicted in Fig.~\ref{fig_method_a} Section A. Our newly proposed decoder architecture is visualised in Fig.~\ref{fig_methods_b} Section B.

\subsection{Architectural building blocks}
\label{subsec:buildingblocks}

$\color{forestgreen}{\mdblksquare}$ \textbf{Convolutional-encoder: }Given a 3D surface patch $\mathcal{S}_{in}\in\mathbb{R}^{D\times H\times W}$ of depth $\bm{D}$, height $\bm{H}$ and width $\bm{W}$, extracted from an automatic multi-label 3D CT segmentation \citep{ts_rsna,isensee_nnu-net_2021} that is separated into binary individual vertebrae contours. We aim to encode the vertebral shape into a low-dimensional embedding $\bm{z}\in\mathbb{R}^M$ with $\bm{M=2048}$ using a fully-convolutional encoder network $f_{conv}(\mathcal{S}_{in})$ parameterised by trainable weights $\bm{w}_{f_{conv}}$. 

$\color{navy}{\mdblksquare}$ \textbf{Point-encoder: } Based on the 3D surface patches  $\mathcal{S}_{in}$, we extract a 3D point cloud $\mathcal{P}_{in}\in\mathbb{R}^{3\times N}$ where $\bm{N}$ corresponds to the number of key-points in the point cloud that are sampled from voxel-representation by applying a threshold on the values in the voxel grid. Using a PointNet \citep{qi2017pointnet} model $f_{point}(\mathcal{P}_{in})$ with weights $\bm{w}_{f_{point}}$, we generate a low-dimensional embedding $\bm{z}\in\mathbb{R}^M$ with $\bm{M=2048}$.

$\color{red}{\mdblksquare}$ \textbf{Graph-encoder: } As for the point-encoder, the graph-encoder utilises an extracted 3D surface point cloud $\mathcal{P}_{in}\in\mathbb{R}^{3\times N}$ for each individual vertebra. We employ a DGCNN \citep{zhang2018end} $f_{graph,k}(\mathcal{P}_{in})$ with parameter $\bm{k}$ for the k-Nearest Neighbour (kNN) graphs and weights $\bm{w}_{f_{graph}}$ to compute an embedding $\bm{z}\in\mathbb{R}^M$ with $\bm{M=2048}$.


$\color{darkslategray}{\mdblksquare}$ \textbf{Convolutional-decoder: }
After generating the embedding $\bm{z}$ a convolutional decoder $g_{conv}(\bm{z})$ with weights $\bm{w}_{g_{conv}}$ is used to map $\bm{z}$ back into $\mathcal{S_{out}}\in\mathbb{R}^{D\times H\times W}$. During training $\mathcal{S}_{out}$ is used to minimise a smooth L1-Loss function $\mathcal{L}_{smoothL1}$ to reconstruct $\mathcal{S}_{in}$. The utilised convolutional decoder model is based on PixelShuffle layers.

$\color{dodgerblue}{\mdblksquare}$ \textbf{Point-decoder: } The aim of the point-decoder $g_{point}(\bm{z})$ is to map $\bm{z}$ back to a 3D Point Cloud representation $\mathcal{P}_{out}\in\mathbb{R}^{3\times N}$ where $\bm{N}$ corresponds to the number of points in $\mathcal{P}_{out}$. Similar to the PixelShuffle layers, this is done by subsequently transferring from network channels $\bm{C}$ into the spatial dimension $\bm{N}$. In the last step $\bm{C}$ is set to 3 to predict $\mathcal{P'}$. During training, we minimise a Loss function based on the Chamfer distance $\mathcal{L}_{chamfer}$ between $\mathcal{P}_{in}$ and $\mathcal{P}_{out}$

$\color{orange}{\mdblksquare}$ \textbf{Point-based shape decoder (pbShape-decoder): } Our point-based shape decoder $g_{pb-shape}(\bm{z})$ can be described as a combination of point decoder and convolutional decoder. In a first step, $\bm{z}$ is mapped to an embedding in shape of key-points $\mathcal{z}_{point}\in\mathbb{R}^{3\times N}$ using a a 3-layer MLP $g_{point-mlp}(\bm{z})$. In a second step, a differentiable extrapolation $g_{sampler}(\bm{z})_{point}$ is performed to map back into a volumetric representation $\mathcal{S}_{out}\mathbb{R}^{D\times H\times W}$. This allows us to then minimise a smooth L1-Loss function $\mathcal{L}_{smoothL1}$ to reconstruct $\mathcal{S}_{in}$ from $\mathcal{S}_{out}$. In contrast to the convolutional-decoder and the point-decoder, our point-based shape-decoder relies on a considerably smaller amount of trainable parameters. A more detailed description of our proposed method is provided in the next section.


\textbf{MLP: } After training of the encoder-decoder models was performed in an unsupervised manner on a large dataset, a small MLP $h(\bm{z})$ is trained on a smaller dataset with fracture labels to predict a binary fracture status from the embedding $\bm{z}$. For this step, the parameters of the previously trained encoder $\bm{w}_{f}$ are fixed, and only the weights of the MLP $\bm{w}_c$ are optimised using a Cross-Entropy Loss function $\mathcal{L}_{BCE}$.

\textbf{Data augmentation: } To improve the generalisability of our models, we introduce affine augmentation on our input data. For encoder models that take a volumetric patch as input, the augmentation is performed before the forward pass during training on each newly drawn batch. For our geometric models, we also apply the affine augmentation on the volumetric patch after drawing the batch and sample a random set of key-points from the augmented volumetric patch using a threshold. 


\begin{table}[!t]
\renewcommand{\arraystretch}{1.3}
\caption{Assessment of architectural building blocks employed within our framework including the number of trainable parameters, total size of the model and computational demands in Mult-Add operations. }
\label{model_details}
\centering
\begin{tabular}{l|c|c|c}
& \textbf{\#Parameter} & \textbf{Total size (MB)} & \textbf{Mult-Adds}\\
\hline
\textbf{Shape encoder: } & & &\\
$\color{forestgreen}{\mdlgblksquare}$
Conv & 6.9M & 69.36 & 5.56G \\
$\color{navy}{\mdlgblksquare}$
Point & 2.8M & 34.1 & 1.39G\\
$\color{red}{\mdlgblksquare}$
Graph & 4.5M & 678.92 & 7.96G\\
\textbf{Shape decoder: } & & &\\
$\color{darkslategray}{\mdlgblksquare}$
Conv & 155k & 24.88 & 1.37G \\
$\color{dodgerblue}{\mdlgblksquare}$
Point & 62k & 0.59 & 0.9M\\
$\color{orange}{\mdlgblksquare}$
Pb- Shape & 62k & 0.66 & 0.68M\\
\textbf{Shape Classifier: } & & &\\
MLP & 541k & 2.18 & 0.54M\\
\end{tabular}
\end{table}

We provide details about the capacity and computational complexity of our AE models 
in Table \ref{model_details}. Our proposed pbShape-decoder requires 600k fewer computations and has only 62k trainable parameters compared to its convolutional counterpart (155k). During the AE-model training, the encoders and decoders are combined, whereas, during training of the MLP classification, only the MLP parameters are adapted. During the test time of the classification, the encoder and MLP parameters are employed.

\subsection{Point-based shape decoder}
\label{subsec:pbshapedecoder}

\begin{figure*}[!t]
\centering
\includegraphics[width=6.3in]{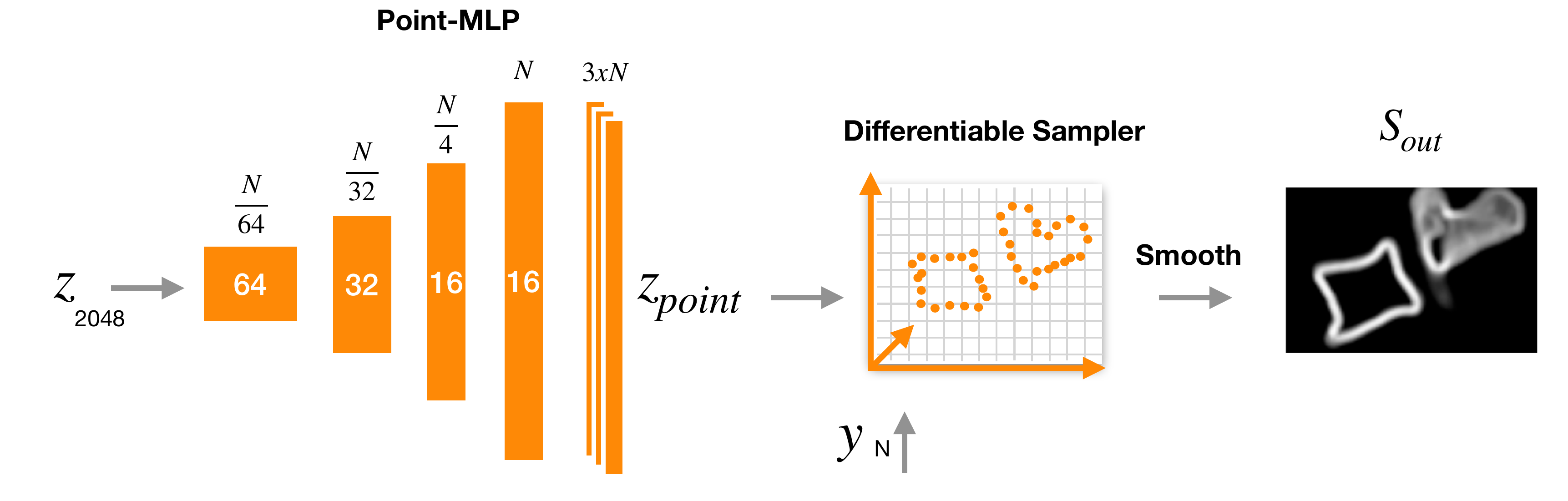}
\caption{\textbf{Point-based Shape decoder: } From the embedding vector $\bm{z}$ a point-representation of $N$ key-points is computed using an MLP. The layers each consist of a 1D-Convolution with the channel size denoted by white font within the blocks, InstanceNorm and ReLU. The number on top of the blocks denotes the size of the spatial dimension.  Afterwards, a differentiable sampling operation is applied on the key-points to obtain a volumetric representation. This step requires $N$ additional parameters $\bm{y}$.}
\label{fig_methods_b}
\end{figure*}

Normally, the decoder $g(\bm{z})$ would follow an inverse architecture comprising transposed 3D convolutions or up-sampling layers and introduce a similar number of additional trainable weights $\bm{w}_{g}$ as the encoder $\bm{w}_{f}$. This makes the latent space less constrained as information about the shape can also be ''encoded" in decoder layers. 
Consequently, we propose a completely new strategy: As opposed to conventional AEs, we map the latent space to represent geometric 3D point coordinates using a small MLP $g_{point-mlp}(\bm{z})$ to compute a point embedding $\mathcal{z}_{point}\in\mathbb{R}^{3\times N}$, which is then used as input to a differentiable sampler $g_{sampler}(\bm{z}_{point})$to reconstruct the original input $\mathcal{S}_{in}$. The trainable weights in $w_{g}$ are limited to $w_{g_{point-mlp}}$ as $w_{g_{sampler}}\rightarrow\varnothing$ (parameters $y$ are fixed after training).
The structure of our proposed decoder is displayed in Fig. \ref{fig_methods_b}.
 
We define $g_{sampler}$ as a differentiable geometric operation that extrapolates a 3D volume from a small number keypoint coordinate and value pairs. Recall that spatial transformer networks \citep{jaderberg2015spatial} perform differentiable interpolation of regular 2D/3D feature maps (or images) given a grid of coordinates that is predicted by a localiser network to obtain sampled values. Each resampled value on an off-grid location depends on 4 (2D) or 8 (3D) neighbouring integer locations that are used within bi-/trilinear interpolation.  Here, following \citep{Heinrich_2023_ICCV} we employ the reverse operation in that spatial coordinates $\bm{x}\in\mathbb{R}^3$ and values $\bm{y}\in\mathbb{R}$ are used as input to extrapolate a 3D grid using their respective trilinear extrapolation coefficients. 

The sampling operation $g_{sampler}$ is implemented as a reverse grid sampling operation in PyTorch, and we compute the derivative for both sampling values $\bm{y}$ and coordinates $\mathcal{z}_{point}$ to make this step differentiable. To mitigate noise and ensure a smooth output, we apply a cubic spline kernel to spatially filter the extrapolation results. Note that the values $\bm{y}$ attached to each key-point are treated as additional trainable parameters, which are shared across all training samples and can be learned restricted within a range of 0 to 1 with a sigmoid. This enables our method to ''paint" a detailed surface, as a zero value placed close to one effectively erases or overwrites parts of a thicker line.

Furthermore, we hypothesise that only $\bm{N}$ representative (key-point) coordinates $\bm{x}$ are required to reconstruct a detailed vertebral surface and that those key-points can be efficiently predicted from the input using $f$ and $g_{point-mlp}$. Hence, our latent space $\bm{z}_{point}$ now represents a geometric entity that reflects a compact representation of the vertebral shape. This is useful for understanding the impact of osteoporotic fractures on vertebral geometry and can hence lead to improved classification accuracy. By keeping the number of trainable parameters $w_{g_{point-mlp}}$ small, we ensure that these advantageous properties of $\bm{z}_{point}$ are mirrored in $\bm{z}$.

\section{Experiments and Results}
\label{sec:experimentsresults}

To explore the efficacy of utilising unsupervised pre-training of vertebral shapes for detecting vertebral body fractures, we conducted two experiments. In the initial experiment, we assessed and compared the classification outcomes of various AE architectures and end-to-end trained models. We also scrutinised the adaptability of segmentation masks generated through deep learning for subsequent diagnostic tasks. The second experiment aimed to determine the robustness of the AE models by executing a data-hold-out experiment on the labelled dataset, specifically during the supervised training phase of the MLP. Subsequently, we outline the prerequisites for our experiments and delve into the implementation details of both the unsupervised pre-training stage for our AE models and the supervised training phase for the MLP. Additionally, we provide implementation details concerning the models subjected to end-to-end training.

\subsection{Datasets and implementation details}
\label{subsec:datasetdetails}

To conduct our experiments as outlined in Fig.\ref{fig_method_a}, we utilise the following two public datasets:

\textbf{TotalSegmentator dataset \citep{ts_rsna,isensee_nnu-net_2021}}: 
The dataset contains 1204 CT scans/patients of different parts of the body and corresponding segmentation masks of 104 volumes of interest including 24 vertebrae (C1-L5). From this dataset, we extracted patches of nearly 13k vertebrae surface masks, which are employed for unsupervised training of our AEs. Using our best classifier model we estimate that 2.9\% are fractured.

\textbf{ VerSe19 dataset \citep{liebl_computed_2021,loffler_vertebral_2020,sekuboyina2021verse}\footnote{(https://osf.io/923ap/, Ethics approval: TUM Proposal 27/19 S-SR, Licence: CC BY-SA 4.0)}: }The dataset is split into 80/40/40 for training, validation and test set and originally serves as a benchmark for the vertebrae segmentation and localisation task. Since fracture labels are also provided for the dataset, we use them to train and test our vertebral fracture classification. The amounts of extracted vertebrae are 770/385/380 respectively. We have also determined that when excluding mild fractures (grade 1) \citep{genant_vertebral_1993}, approximately 10\% of vertebrae of each subset are fractured.

\textbf{Unsupervised Pre-Training}: The training of our AE models is performed using the same training regime and hyper-parameters for all encoder-decoder combinations, to ensure comparability. Each model is trained for 15000 iterations using a batch size of 64. The models are optimised by deploying the Adam optimiser with an initial learning rate of 0.005 and CosineAnnealing scheduler with two warm restarts. During training, we augment our data by applying affine transformations. For geometric approaches, we have fixed the number of points to 3840, as we have found that the shape of vertebrae depicted in our given patch size can be accurately represented with this amount of key-points. For the DGCNN (graph-encoder) model, we have fixed the hyper-parameter for graph neighbours to $k=20$.

\textbf{Supervised Training}: For our MLP-Classifiers, the training is conducted for 6000 iterations using batch-size 16. We again use the Adam optimiser with an initial learning rate of 0.0001 and CosineAnnealing learning rate scheduling with warm restarts. To enhance the diversity of our input data to the classification model, we augment the input patch to the encoder using affine transformations and we also keep the parameters of the encoder fixed. For geometric encoder models, we utilise the augmented input to construct a randomly sampled point cloud of the vertebral surface. During test time, this point cloud is sub-sampled by employing farthest point sampling to determine 3840 surface points instead of drawing random points from an intensity threshold. While we use the ground truth (GT) segmentation masks for training our classifier models, we employ automatically generated segmentation masks using the TotalSegmentator model \citep{ts_rsna, isensee_nnu-net_2021} at test time.

\textbf{End-to-End Training}: 
In addition to our AE architecture comparison, we train our encoder models from scratch, namely the convolutional encoder, the PointNet and the DGCNN, in combination with the MLP classifier in an end-to-end fashion on the Verse19 dataset. Furthermore, the convolutional encoder is trained on image-patches containing the intensities values as information as well as on surface patches. 




\subsection{Automated fracture detection pipeline}
\label{subsec:fxpipeline}

We evaluate the accuracy of our shape-based vertebral fracture detection through a binary classification that distinguishes between healthy and fractured vertebrae. Since mild fractures (Genant grade 1)\citep{genant_vertebral_1993} of the vertebral bodies are oftentimes indistinct and thus harder to annotate consistently (higher inter-rater variability), vertebrae with label fracture grade 1 are left out for this analysis.

\begin{table}[!t]
\renewcommand{\arraystretch}{1.3}
\caption{Median AUC on the VerSe19 test set over 10 seeds}
\label{tab:results}
\centering
\begin{tabular}{l||c|c}
 & GT-masks & TS-masks \\
Model &  &  \\
\hline
\textbf{End-to-end} &  & \\
Conv-encoder (img) $\color{forestgreen}{\mdlgblksquare}$ & 0.756 & - \\
Conv-encoder (surf) $\color{forestgreen}{\mdlgblksquare}$ & 0.883 & 0.847\\
Point-encoder $\color{navy}{\mdlgblksquare}$ & 0.842 & 0.803 \\
Graph-encoder $\color{red}{\mdlgblksquare}$ & 0.615 & 0.634\\
\hline
\textbf{Conv-encoder -} & & \\
Conv-decoder 
$\color{forestgreen}{\mdlgblksquare}$-$\color{darkslategray}{\mdlgblksquare}$
& 0.890 & 0.814\\
Point-decoder 
$\color{forestgreen}{\mdlgblksquare}$-$\color{dodgerblue}{\mdlgblksquare}$
& 0.704 & 0.712\\
pbShape-decoder
$\color{forestgreen}{\mdlgblksquare}$-$\color{orange}{\mdlgblksquare}$ 
& 0.924 & \textbf{0.914}\\
\hline
\textbf{Point-encoder -} & & \\
Conv-decoder
$\color{navy}{\mdlgblksquare}$-$\color{darkslategray}{\mdlgblksquare}$ 
& 0.930 & 0.882\\
Point-decoder 
$\color{navy}{\mdlgblksquare}$-$\color{dodgerblue}{\mdlgblksquare}$
& 0.911 & 0.859\\
pbShape-decoder 
$\color{navy}{\mdlgblksquare}$-$\color{orange}{\mdlgblksquare}$
& \textbf{0.937} & 0.901\\
\hline
\textbf{Graph-encoder -} & & \\
Conv-decoder
$\color{red}{\mdlgblksquare}$-$\color{darkslategray}{\mdlgblksquare}$
& 0.871 & 0.825\\
Point-decoder
$\color{red}{\mdlgblksquare}$-$\color{dodgerblue}{\mdlgblksquare}$ 
& 0.843 & 0.760\\
pbShape-decoder
$\color{red}{\mdlgblksquare}$-$\color{orange}{\mdlgblksquare}$ 
& 0.920 & 0.856\\

\end{tabular}
\end{table}

In Table \ref{tab:results} we show the median Area-Under-Curve (AUC) of fracture detection results for 10 seeds of training the MLP for various encoder-decoder architectures and additional end-to-end trained models on the test set. We report the results both for using ground truth segmentation masks for preprocessing and for automatically generated segmentation masks using the TotalSegmentator \citep{ts_rsna}\citep{isensee_nnu-net_2021}.

The results illustrate the benefits of pre-training on shape reconstruction for the detection of vertebral fractures. Furthermore, the results convey that explicitly training on the shape of vertebrae improves classification performance compared to training on image intensities directly. The highest AUC score in the setup utilising GT segmentation masks to generate the model input is achieved with the Point-encoder and pbShape decoder architecture, resulting in an AUC of 0.937. When employing TotalSegmentator (TS) segmentation masks, the convolutional-encoder in conjunction with the pbShape decoder attains an AUC of 0.914
Notice that the pbShape-decoder consistently achieves the highest AUC for each encoder model, highlighting the benefits of the robustness of our proposed decoder. 

When examining the outcomes between AE models and end-to-end trained models, the AE models consistently produce more accurate and robust classification results.
Especially employing the same convolutional-encoder architecture, outperforms its image-based end-to-end trained counterpart by more than 12.8\% in AUC. Even when using segmentation masks from TotalSegmentator, there's a 9\% improvement. 

Another noteworthy observation arises when considering realistic segmentation errors by adapting TS-segmentation labels, especially in models trained with graph-based encoding a more severe drop in classification performance can be discerned in contrast to other encoder models.

\begin{figure*}
    \centering
    \includegraphics[width=2.1in]{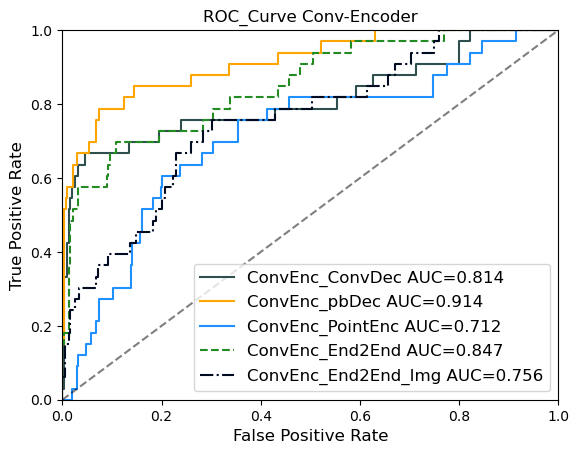}
    \includegraphics[width=2.1in]{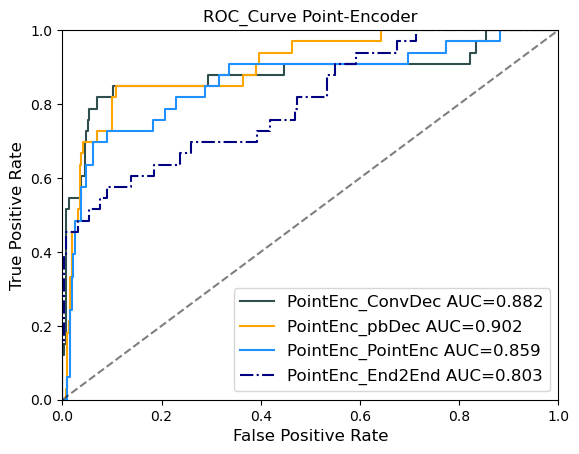}
    \includegraphics[width=2.1in]{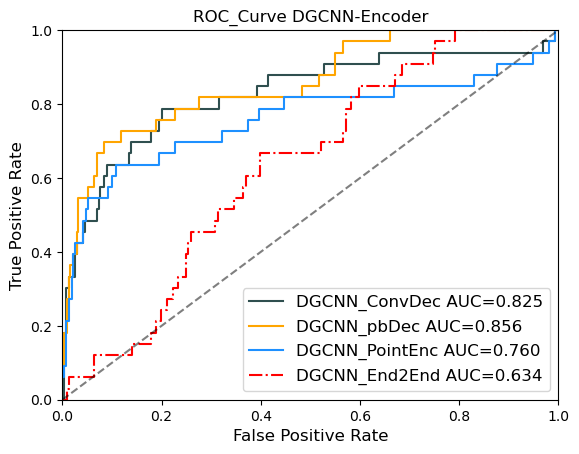}
    \caption{ROC-Curve and corresponding AUC for encoder-decoder combinations of the median AUC of 10 seeds. The encoders are grouped by colour and linestyle, whereas the decoders are grouped by colour and marker. The corresponding Area Under Curve (AUC) is listed inside the legend.}
    \label{fig:roc_fx}
\end{figure*}

To investigate the performance of our models in greater detail, we present the receiver-operator curves (ROC-curves) in Fig. \ref{fig:roc_fx}. The depicted results are computed using TS-generated segmentation masks for preprocessing. From left to right the ROC-curves are provided for convolutional-encoder, point-encoder and graph-encoder models.

When interpreting the plot for the convolutional encoder, identifying the threshold with the highest possible sensitivity and specificity would be in favour of the ConvEnc-pbDec model which also reaches the highest overall AUC in this setup (AUC=0.914). 

For ROC-curves obtained for both point-encoder and graph-encoder models, we observe that by choosing an optimal threshold either the convolutional decoder model or the pbShape decoder produces the most accurate classification outcome. 

\subsection{Data-hold-out experiment}
\label{subsec:dataholdexperiment}

\begin{figure*}
    \centering
    \includegraphics[width=2.8in]{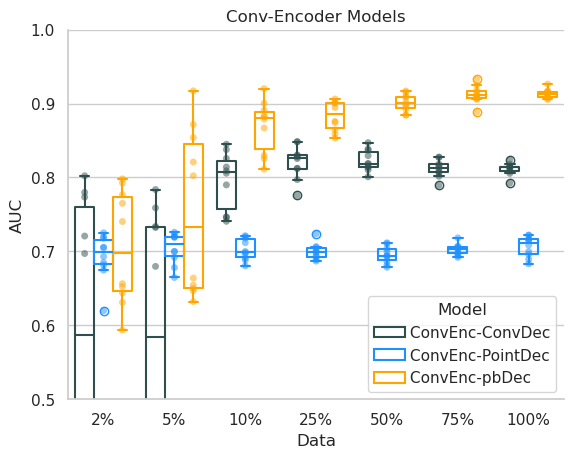}
    \includegraphics[width=2.8in]{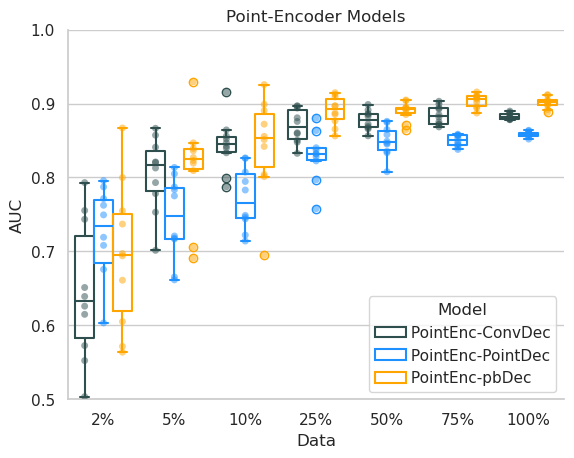}
    \includegraphics[width=2.8in]{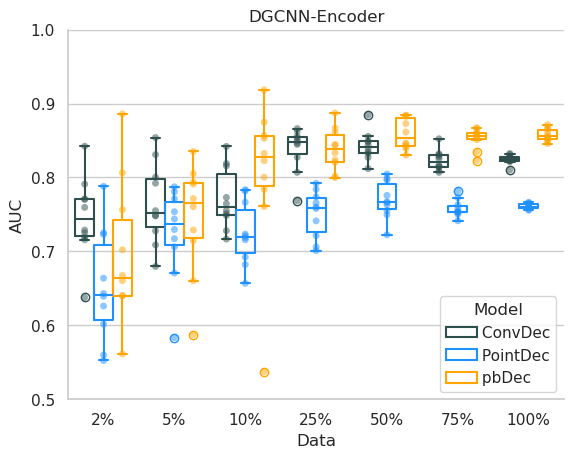}
    \includegraphics[width=2.8in]{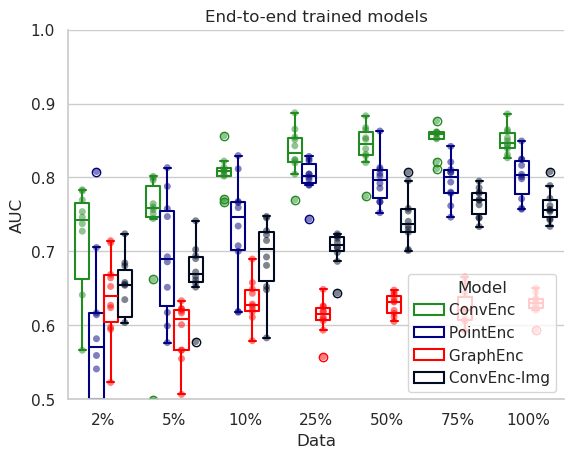}
    \caption{Results of our data-hold-out experiment as boxplots and scatterplots of the AUC obtained for 10 random seeds each. The plots are separated by the employed encoder architecture, and provide the classification results obtained with the respective decoder. \textbf{Top-Left: } Convolutional-encoder \textbf{Top-Right: } Point-encoder models, \textbf{Bottom-Left: } Graph-encoder models \textbf{Bottom-Right: } End-to-End trained models including the traditional CNN trained on image intensities instead and trained on vertebra surface (denoted as Conv-encoder img and surf).}
    \label{fig:datahold}
\end{figure*}

In our data-hold-out experiment, we explore the required amount of supervised training data for achieving robust classification outcomes. AUC values, calculated across 10 random seeds for each data split (2\%, 5\%, 10\%, 50\%, 75\%, and 100\%), are presented as both scatter plots and box plots over the 10 seeds (see Fig. \ref{fig:datahold}). The experiment demonstrates that the robustness is increased by adding more training data. Notably, employing just 25\% of training data already yields as accurate and robust classification results as models trained on the complete training dataset. 
The increased number of outliers in experiments involving training data below 25\% indicates that specific data splits offer more advantageous conditions for the classification task, while others present less favourable circumstances.

\section{Discussion}
\label{sec:discussion}

In this work, we examined the benefits of directly utilising shape information in the diagnosis of vertebral body fractures. We introduced a shape encoding approach that takes a volumetric surface representation of a vertebra as input, encoding it into a latent vector, which is decoded by transforming the latent vector into point representations (using an MLP) and then applying a novel differentiable point sampling method: the Point-based Shape decoder (pbShape-decoder). 

Within our vertebrae AE framework, we analyse the performance of our proposed decoder and other AE building blocks that are pre-trained on a large-scale dataset to reconstruct the shape of an input vertebra surface patch and additionally employ end-to-end trained models for comparison. Whereas the vast majority of weights in our AE models (the whole encoder part) are trained in an unsupervised fashion, only a light-weight MLP classifier using the sparse shape representation from the latent space requires training with fracture labels. 
The findings of our end-to-end trained models indicate that, with supervised training on surface data, good classification results (AUC above 0.9) can already be obtained using a 3D CNN model and our results demonstrate that using automatically generated segmentation data yields preferable results in comparison to training on image-intensity data directly. This shows the advantage of explicitly employing shape information from large-scale multi-label segmentation models over implicitly learning shape information in intensity-based CNNs.
Furthermore, employing surface information that was previously already learned, for instance by the nnUNet in the TotalSegmentator model\citep{ts_rsna,isensee_nnu-net_2021} enhances the robustness of the classification task against domain shifts. The results affirm that the quality and robustness of the generated TotalSegmentator \citep{ts_rsna,isensee_nnu-net_2021} masks are adequate for this task. Furthermore, our findings highlight that the choice of architecture influences the model's robustness in the face of inaccuracies in segmentation masks. 
In a clinical setting, obtaining ground truth segmentation masks can be challenging. The fact that our model achieves an AUC of 0.914, holds significant potential for automated diagnosis of osteoporotic fractures in the spine with limited availability of labelled data.

Our findings further suggest that, when trained in an end-to-end fashion, geometric learning approaches such as PointNet \citep{qi2017pointnet} and DGCNN \citep{zhang2018end} do not perform as well as 3D-convolutional approaches. However, when integrated as an encoder into an AE architecture and pre-trained on a large-scale dataset, we demonstrate their ability to achieve more accurate results. Moreover, our findings indicate that models employing a PointNet as an encoder produce accurate and robust results with only a small deviation between the different decoder models (AUC pbShape: .901 conv: -1.9\% and point: -4.2\%) with fewer trainable parameters being incorporated. This underscores the effectiveness and efficacy of the PointNet architecture.
Contrarily, the graph-based encoder, appears to under-perform on this specific task. This could be attributed to the nature of the task of detecting vertebral fractures which involves a more global geometric relationship based on height differences, rendering local graph structures less relevant. Consequently, capturing finer geometric features may not contribute significantly to this task but might be of interest when detecting other degenerative deformations of vertebrae.

By performing a data-hold-out experiment, we highlight the effectiveness and robustness of our models and find that an AUC above 0.9 can be obtained even when training on 5\% of data (51 healthy and 5 fractured vertebrae), but robustness increases when training on a larger dataset. 
The results of this experiment furthermore show the importance and impact of data distributions and the prevalence of vertebral fractures for this task increases the difficulty to compare methods between multiple datasets. We thus consider our point-encoder point-decoder model similar to the approach proposed as pAE by Sekuboyina et al. \citep{sekuboyina_probabilistic_2019}, who also utilise a point cloud generated from ground truth segmentation masks as input into their point-based AE architecture. However, they treat the detection of vertebral body fracture as an anomaly detection task by training the reconstruction of the point cloud on healthy samples only achieving an AUC of 0.759 with the best model on their dataset using accurate segmentation masks. In comparison, our purely point-based AE model reaches an AUC of 0.895 (0.911 on ground truth segmentation masks), setting a new state-of-the-art.

\section{Conclusion}
\label{sec:conclusion}
In summary, our study explores AE architectures for effective shape encoding of vertebrae to diagnose osteoporotic fractures of vertebrae. We introduce a novel approach, utilising a differentiable geometrical decoder for unsupervised pre-training on the TotalSegmentator dataset. By combining convolutional or point-based encoders with our proposed shape decoder, we achieve a meaningful and robust shape representation in the latent space, facilitating the detection of osteoporotic fractures. Our approach demonstrates robustness through comparisons with ground truth segmentation masks, showcasing commendable classification results even when applying automatically generated segmentation masks. Moreover, this approach can be extended to other shape-related tasks, e.g. diagnosing degenerative deformations, spinal canal stenosis or lymph nodes.

\section*{Acknowledgment}

This work was financially supported by the German Federal Ministry of Education and Research under Grant: 01EC1908D.

\bibliographystyle{unsrtnat}
\bibliography{shape_matters}

\end{document}